\documentclass[baaa]{baaa}

\usepackage[pdftex]{hyperref}
\usepackage{subfigure}
\usepackage{natbib}
\usepackage{helvet,soul}
\usepackage[font=small]{caption}

\contriblanguage{1}

\contribtype{1}

\thematicarea{1}

\title{Evolution of the colour-magnitude relation of early-type
galaxies in cosmological numerical simulations}

\titlerunning{Evolution of the CMR of early-type galaxies}

\author{L.J. Zenocratti\inst{1,2}, A.V. Smith Castelli\inst{1,2}, M.E. De Rossi\inst{3,4}, \& F.R. Faifer\inst{1,2}}
\authorrunning{Zenocratti et al.}

\contact{lzenocratti@fcaglp.unlp.edu.ar}

\institute{Facultad de Ciencias Astron\'omicas y Geof\'isicas, UNLP, Argentina \and
Instituto de Astrof\'isica de La Plata, CONICET--UNLP, Argentina \and
Ciclo B\'asico Com\'un, UBA, Argentina \and
Instituto de Astronom\'ia y F\'isica del Espacio, CONICET--UBA, Argentina}

\abstract{
In this work, we study the evolution with redshift of the colour-magnitude relation (CMR) of early-type galaxies. This evolution is analyzed through cosmological numerical simulations from $z=2$ to $z=0$. The preliminary results shown here represent the starting point of a study aimed at identifying the processes that originated the observed CMR of early-type galaxies at z = 0.
}

\resumen{
En este trabajo estudiamos la evoluci\'on con el {\em redshift} de la relaci\'on color-magnitud (CMR) de galaxias de tipo temprano. Esta evoluci\'on es analizada a partir de simulaciones num\'ericas cosmol\'ogicas, desde $z=2$ hasta $z=0$. Los resultados preliminares mostrados aqu\'i representan el punto de partida de un estudio apuntado a identificar los procesos que originaron la CMR de galaxias de tipo temprano observada a $z=0$.
}


\begin{document}

\maketitle

\section{Introduction}
\label{S_intro}

In the colour-magnitude diagram (CMD), early-type (ET) galaxies trace a well-defined sequence from dwarfs to giants, where brighter galaxies tend to be redder (e.g., \citealp{chen_2010}; \citealp{smith_2013}; \citealp{roediger_2017}). This colour-magnitude relation (CMR) is considered as universal, in the sense that it is observed with similar slopes in rich groups and clusters in the local Universe ($z\sim 0$). This sequence is interpreted as a mass-metallicity relation, with brighter and redder galaxies tending to be more massive and more metal-enriched (\citealp{sanchez_2006}; \citealp{conroy_2014}). Nevertheless, processes that establish and define it are not yet totally known for certain (e.g., \citealp{roediger_2011}; \citealp{janz_2017}; \citealp{connor_2019}). 

In this work, we present a preliminary study that extends a previous analysis of the CMD for ET galaxies extracted from cosmological numerical simulations, studying the behaviour of the diagram as function of redshift. Our main goal is to provide clues that explain the origin of the CMR through the study of its evolution since the formation times of these galaxies until today. 

\section{Simulated galaxies} \label{simulation}

\subsection{The EAGLE simulations}

In this work, we use simulations of the EAGLE (Evolution and Assembly of GaLaxies and their Environments) project (\citealp{schaye_2015}; \citealp{mcalpine_2016}), a suite of cosmological, hydrodynamical simulations of a standard $\Lambda$CDM universe, that were performed by using a modified version of the GADGET-3 code (\citealp{springel_2005}; \citealp{schaller_2015}). The cosmological parameters used for the EAGLE simulations are those of the Planck Collaboration (\citealp{planck_2014}): $\Omega_\Lambda=0.693$, $\Omega_m=0.307$, $\Omega_b=0.048$ and $h=0.6777$. We started working with the reference, intermediate-resolution simulation Ref-L0100N1504, which has a box size of $100~\rm{cMpc}$, with an initial baryonic particle mass of $1.81\times10^6~\rm{M}_\odot$ and a maximum proper softening length of $0.70~\rm{pkpc}$.

The identification of galaxies in the simulations were carried out by applying a Friends-of-Friends technique (\citealp{davis_1985}), combined with the SUBFIND algorithm (\citealp{springel_2001}; \citealp{dolag_2009}). The subgrid physics in EAGLE simulations implement prescriptions for radiative cooling and heating, star formation, chemical enrichment, supernovae and active galactic nuclei feedbacks, and interactions and fusions, among other processes (see \citealp{schaye_2015} for details).

\begin{figure*}[!ht]
  \centering
  \includegraphics[width=0.4\textwidth]{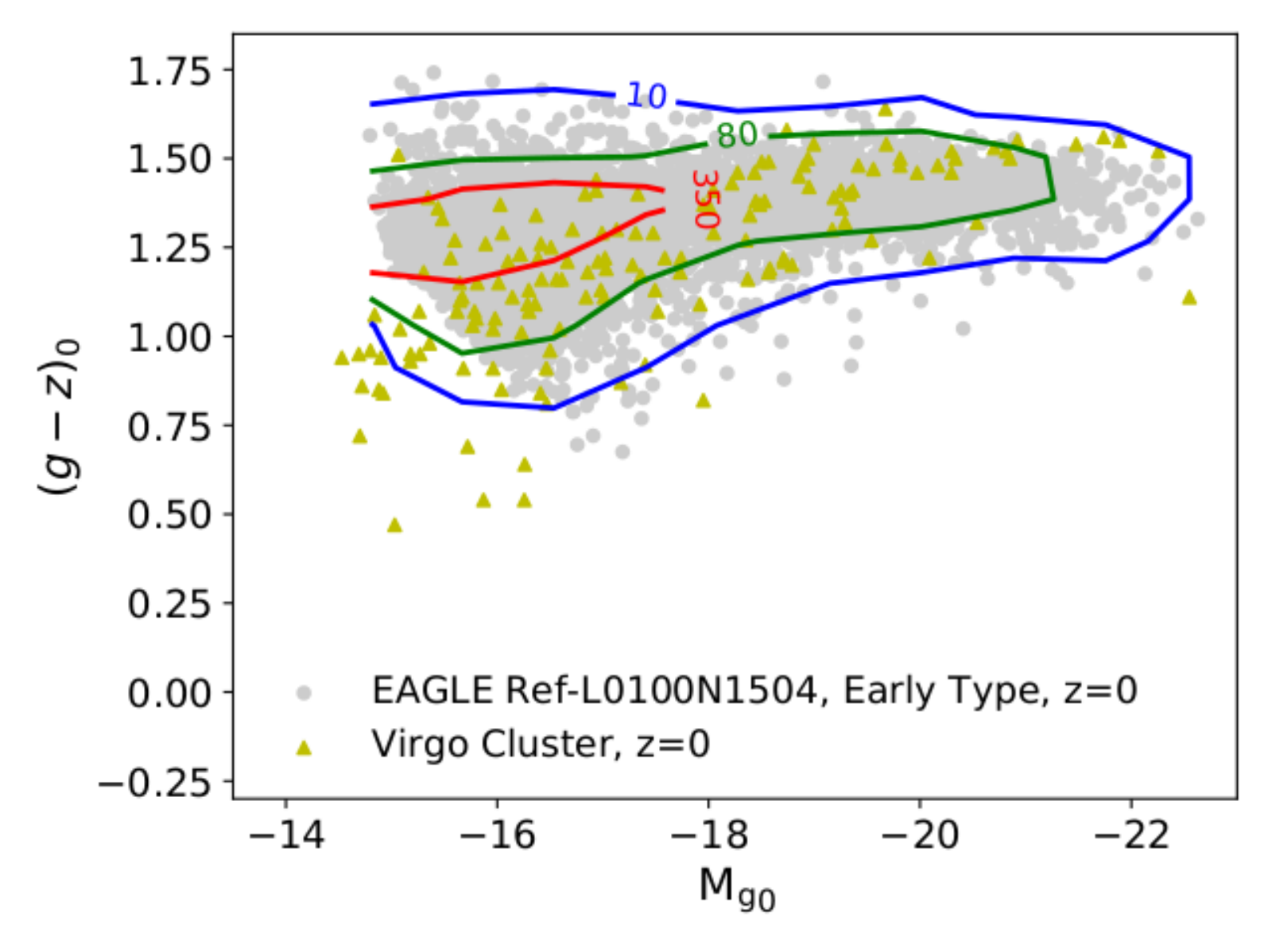}
  \includegraphics[width=0.4\textwidth]{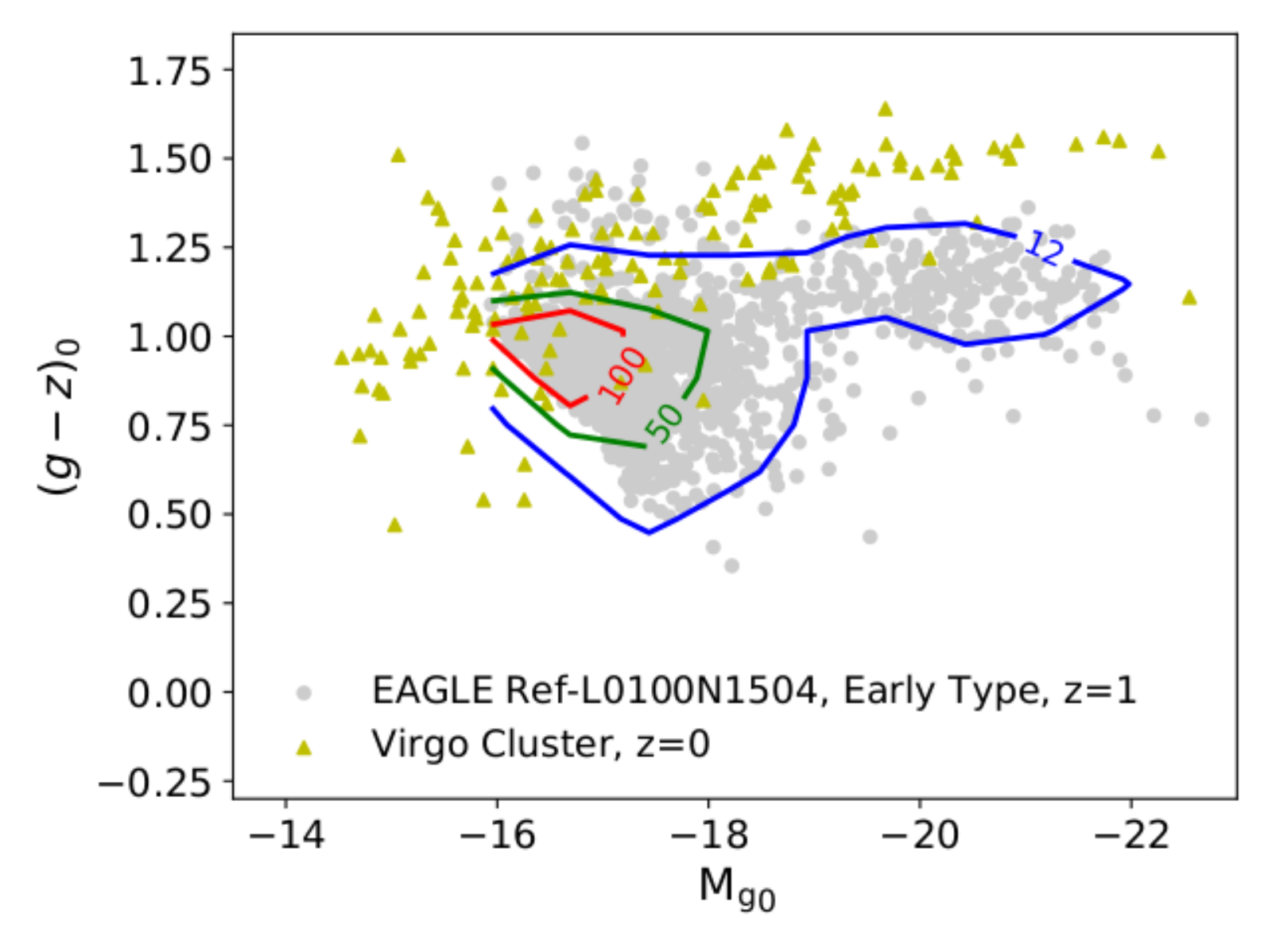}
  \par
  \includegraphics[width=0.4\textwidth]{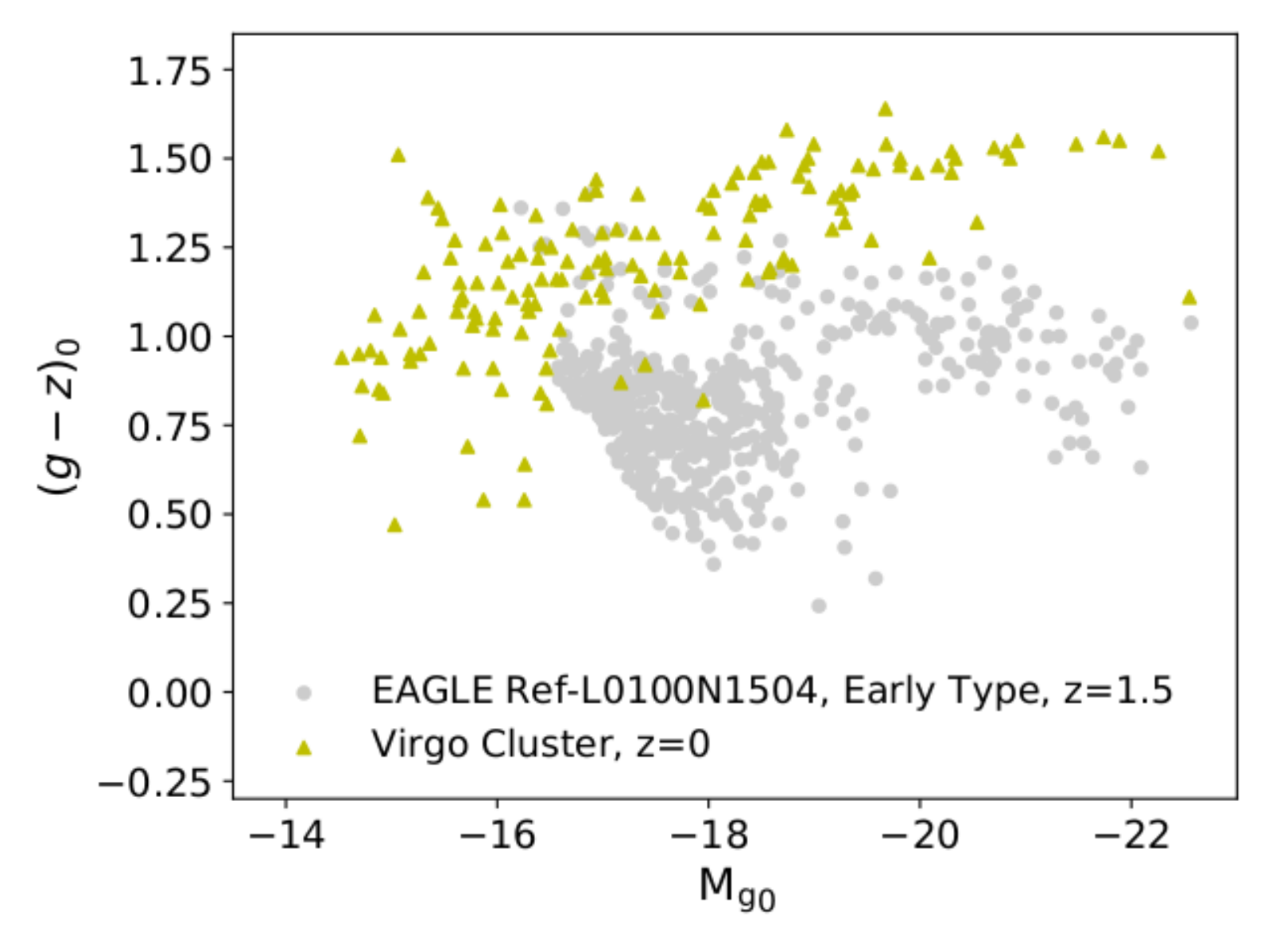}
  \includegraphics[width=0.4\textwidth]{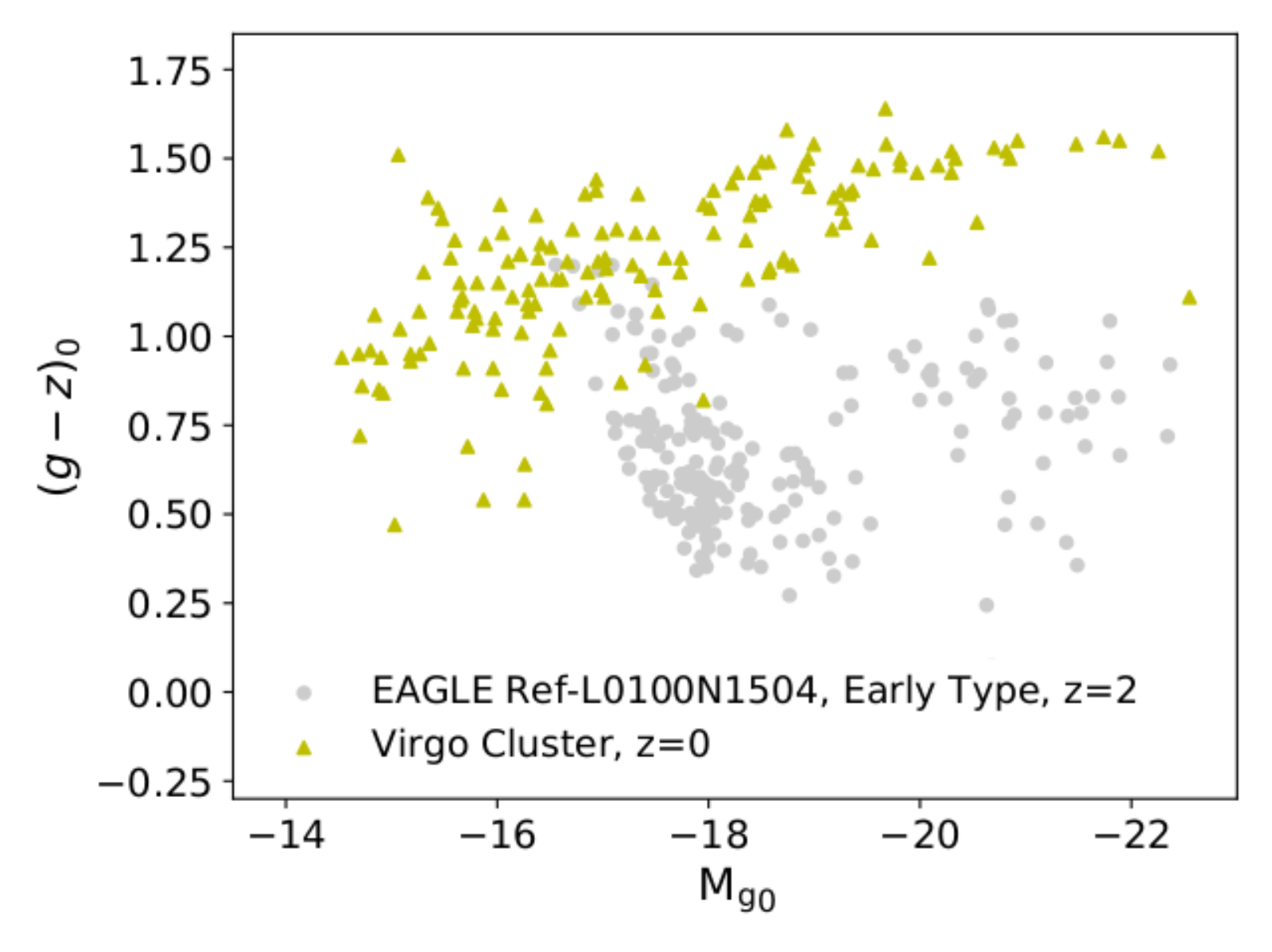}
  \caption{Colour-magnitude diagrams of early-type galaxies at different $z$. Grey circles are simulated galaxies, extracted from the EAGLE Ref-L0100N1504 simulation using the selection criteria stated in Section \ref{selection}. Yellow triangles are galaxies of the Virgo Cluster (\citealp{chen_2010}; \citealp{sanchezjanssen_2019}). Different panels show the CMD at $z=0$ (top left), $z=1$ (top right), $z=1.5$ (bottom left), and $z=2$ (bottom right). Three contours of constant number of galaxies are shown for $z=0$ and $z=1$.}
  \label{fig1}
\end{figure*}

\subsection{Galaxy selection}
\label{selection}

From the Ref-L0100N1504 EAGLE simulation, we extracted galaxies with stellar mass $M_\star\geqslant10^9~\rm{M}_\odot$, a star formation rate ($SFR$) such that $\log(SFR/M_\star)\leqslant -11~\rm{yr}^{-1}$, and a star forming gas fraction $M_{SF\ gas}/(M_{SF\ gas}+M_\star)\leqslant 0.1$ (\citealp{zenocratti_2018}). Galaxies fulfilling these conditions are defined as our simulated sample of ET galaxies. At all analysed redshifts, our sample of selected galaxies is constituted by more than $9\,500$ systems with stellar masses higher than $10^9~\rm{M}_\odot$. To check our selection criteria, we compared the CMD of that sample at $z=0$ with the corresponding to ET galaxies in the Virgo Cluster (\citealp{chen_2010}; \citealp{sanchezjanssen_2019}). The CMD of the simulated sample at $z=0$ agrees with the observed one (see Fig. \ref{fig1}).

\section{Results}
\label{results}
\subsection{CMR at different $z$}

Panels of Fig. \ref{fig1} show the simulated $(g-z)_0$ vs. $\rm{M}_{g0}$ CMD from $z=0$ to $z=2$ (grey circles); for comparison, galaxies of the Virgo Cluster ($z\approx0$) are shown in every panel (yellow triangles). As can be seen, the simulated sample and the observed one at $z=0$ are located in the same region of this CMD, therefore our selection criteria lead to a sample of simulated galaxies which follows the observed trends. Also, some contours of constant number of simulated galaxies are plotted in the first two panels (i.e., at $z=0$ and $z=1$), in order to identify which region of the $(g-z)_0$ vs. $\rm{M}_{g0}$ plane is more densely populated. As can be seen, when $z$ decreases the number of ET galaxies increases in both luminosity extremes. That increment seems to be higher in the less luminous region of the CMD. Also, at lower $z$ the bulk of galaxies in the CMD is located in redder regions. Finally, it can be seen that the brightest region of the CMD is always populated.   

\subsection{Distributions of magnitudes and colours at different $z$}

\begin{figure}[!ht]
  \centering
  \includegraphics[width=0.4\textwidth]{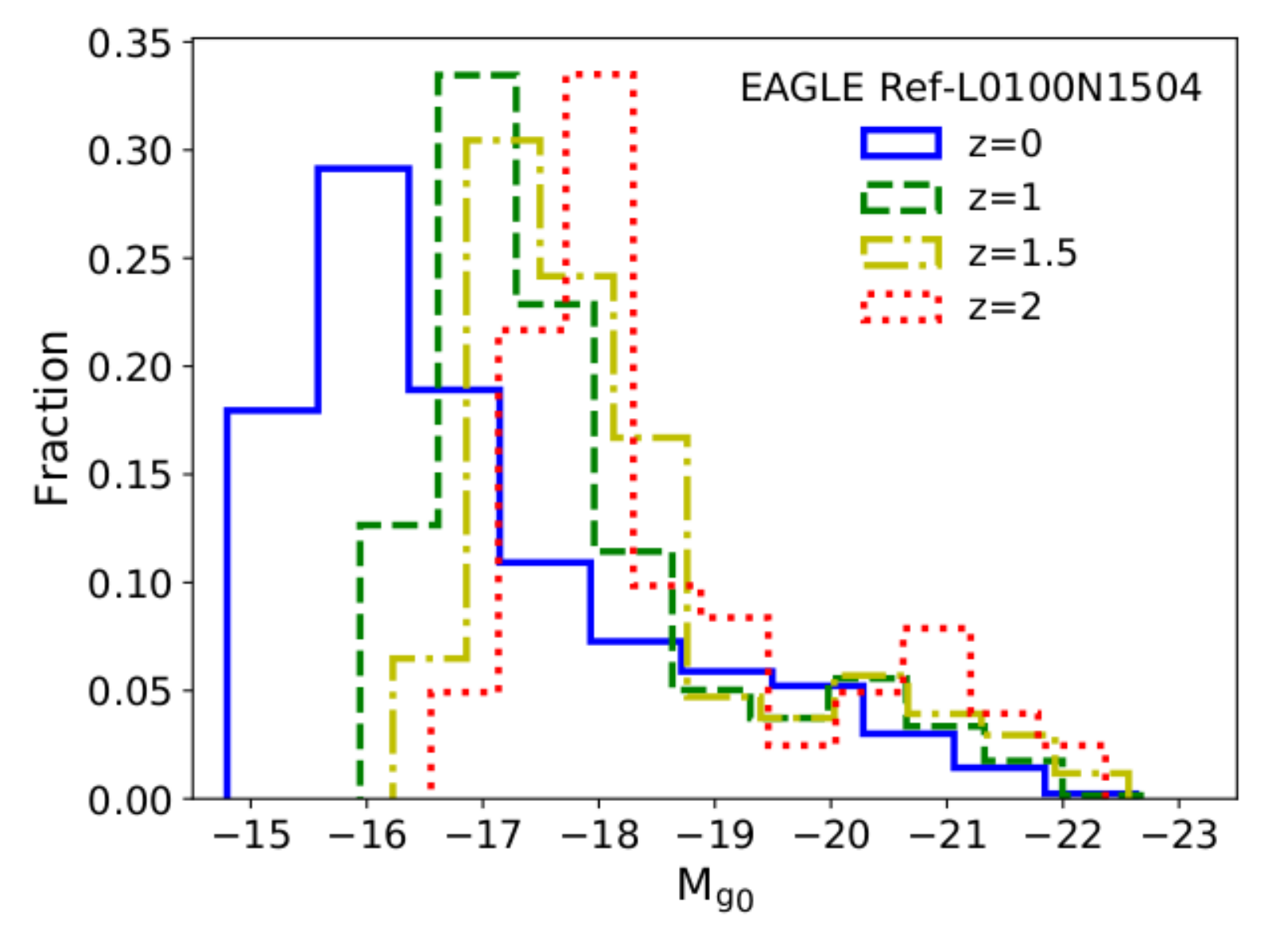}
  \includegraphics[width=0.4\textwidth]{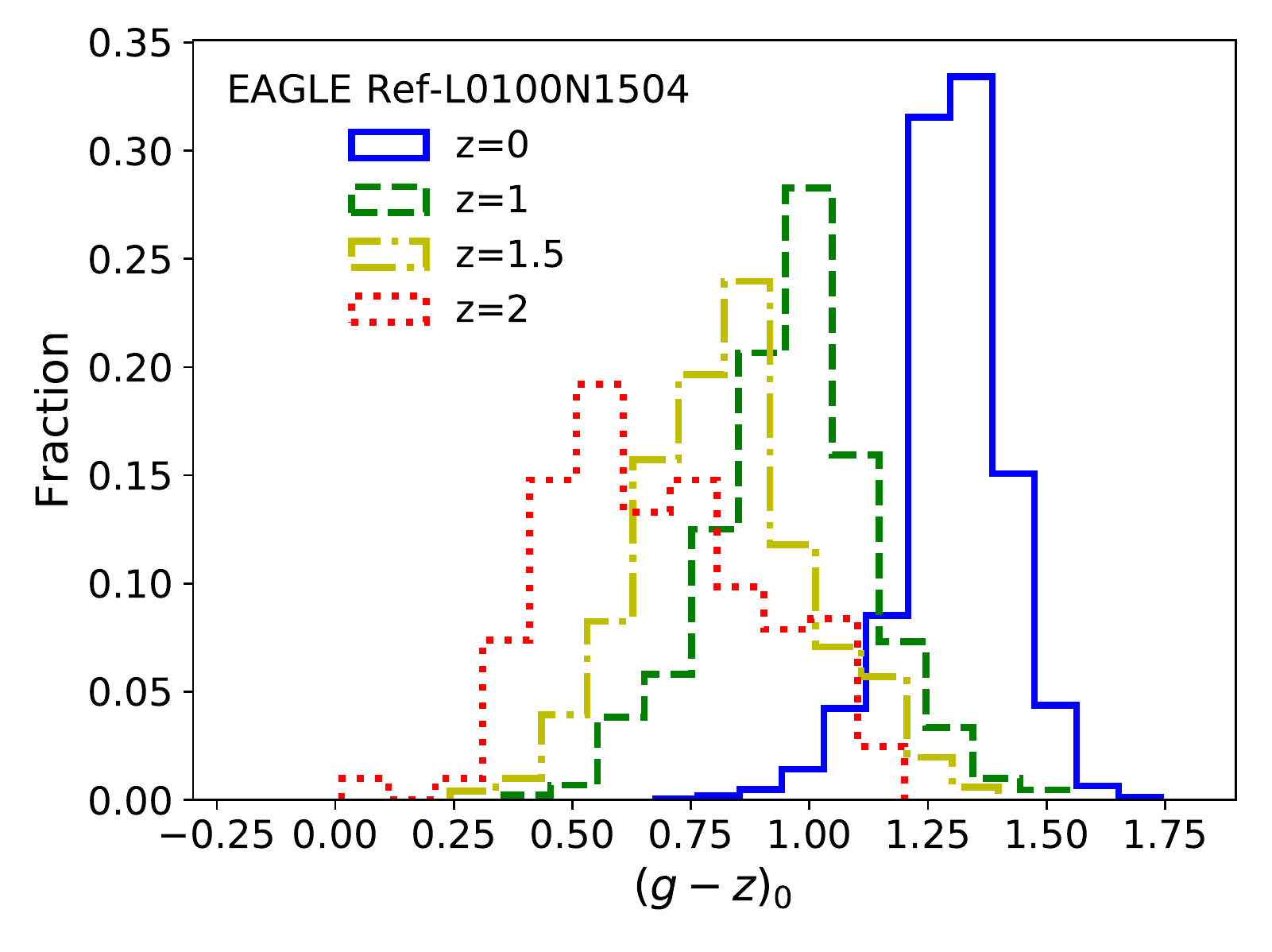}
  \caption{Distribution of $M_{g0}$ magnitude (top) and $(g-z)_0$ colour (bottom) at different redshifts $z$ for the simulated sample of ET galaxies, extracted from the EAGLE Ref-L0100N1504 simulation. The histograms correspond to $z=0$ (blue solid), $z=1$ (green dashed), $z=1.5$ (yellow dot-dashed), and $z=2$ (red dotted).}
  \label{fig2}
\end{figure}

The panels of Fig. \ref{fig2} show the distributions of magnitude $\rm{M}_{g0}$ and colour $(g-z)_0$ for redshifts from $z=0$ to $z=2$, for the simulated sample of ET galaxies. With respect to the distribution in magnitude, the low-magnitude peak moves towards less luminous magnitudes at lower $z$. Also, at $z=2$ a bimodality can be seen, with a peak at $M_{g0}\approx-18$ and another (much less prominent) peak at $M_{g0}\approx-21$. As $z$ decreases, the brightest peak vanishes and the faintest galaxies become the dominant population, which is in agreement with observations: at z=0 faint ET galaxies are much more abundant than bright ET galaxies. In addition, as $z$ decreases, the faint peak of the luminosity distribution moves towards faintest magnitudes showing that the faintest end of the CMR is the last region to be populated in the CMD, within our considered mass range ($M_\star>10^9~M_\odot$).
 
The distribution of colour $(g-z)_0$ shows that at lower $z$, the peak of the distribution moves towards redder colours, being this distribution narrower, i.e., the scatter in colour seems to decrease towards lower $z$. This means that simulated ET galaxies (selected with our criteria) tend to be bluer in the past, showing colours within a little broader range than observed today. In particular, the bluest galaxies at high redshift might be ET galaxies going through their final stages of star formation. The shift of the maximum with $z$ towards redder colours might be due to an increase in the average age, while the lower scatter towards $z=0$ might indicate that $(g-z)_0$ is not sensitive to age variations within the population. Specific properties of our simulated sample (such as masses, ages, SFRs, metallicities, among others) will be studied exhaustively in a forthcoming work.


\section{Conclusions and further work}
\label{summary}

We started studying the evolution with redshift of the CMR for simulated ET galaxies, extracted from the EAGLE Ref-L0100N1504 simulation. Our selected sample of simulated $z=0$ galaxies is consistent with the observed CMD for ET galaxies in the Virgo Cluster. Therefore, we also used these criteria to select ET galaxies up to $z=2$. In the CMD, ET galaxies are placed in bluer regions with increasing $z$. At lower $z$, the number of systems with lower luminosities increases, and ET galaxies tend to be redder, with a decreasing scatter in the colour distribution. Blue and bright ET galaxies present at higher $z$ might be systems still in development.

In a future work, additional properties of the selected sample of ET galaxies (mass, metallicities, SFRs, etc.) will be studied in detail. The evolution of such properties will be analysed comprehensively, as well as the evolution of dynamics and kinematics in these systems, aiming at better understand the setting of the CMR up until the present. Also, other EAGLE simulations with variations in the subgrid physics will be used, in order to determine how variations in parameters affect the CMR.

\begin{acknowledgement}
We acknowledge Asociaci\'on Argentina de Astronom\'ia for giving us the space to show our results. We acknowledge support from PICT-2015-3125 of ANPCyT, PIP 112-201501-00447 of CONICET and G151 of UNLP (Argentina). We acknowledge the Virgo Consortium for making
their simulation data available. The EAGLE simulations were performed using the DiRAC-2 facility at Durham, managed by the ICC, and the PRACE facility Curie based in France at TGCC, CEA, Bruy\`{e}resle-Ch\^{a}tel.
\end{acknowledgement}


\bibliographystyle{baaa}
\small
\bibliography{biblio_zenocratti}
 
\end{document}